\documentclass[aps,showpacs,amsmath,amssymb,nofootinbib]{revtex4}
\usepackage{graphicx}
\usepackage{amssymb}
\newcommand{\mb}[1]{\mbox{\scriptsize #1}}
\newcommand{\sigmac}{\bar{q}\tau^{i}q}
\newcommand{\pic}{\bar{q}i\gamma_5\tau^{i}q}

\newcommand{\rhom}{\rho_{\mb{MRE}}}
\newcommand{\mre}{\int_{\mb{MRE}}}
\newcommand{\bvec}[1]{\mbox{\boldmath $#1$}}

\begin{document}
\title{Stability of Color-Flavor Locked Strangelets}
\author{O. Kiriyama}
\email{kiriyama@th.physik.uni-frankfrt.de}
\affiliation{Institut f\"ur Theoretische Physik, J.W. Goethe-Universit\"at, 
D-60439 Frankfurt am Main, Germany}
\date{\today}

\begin{abstract}
The stability of color-flavor locked (CFL) strangelets is studied 
in the three-flavor Nambu--Jona-Lasinio model. 
We consider all quark flavors to be massless, for simplicity. 
By making use of the multiple reflection expansion, 
we explicitly take into account finite size effects 
and formulate the thermodynamic potential for CFL strangelets. 
We find that the CFL gap could be large enough so that 
the energy per baryon number of CFL strangelets is greatly affected. 
In addtion, if the quark-quark coupling constant is larger than 
a certain critical value, there is a possibility 
of finding absolutely stable CFL strangelets.
\end{abstract}
\maketitle

\section{Introduction}
The properties of quark matter 
have attracted a good deal of interest 
since it has been suggested that strange quark matter 
could be a ground state of strongly interacting matter.\cite{EW,SQM} 
The most likely place for strange quark matter, perhaps, is 
the interior of compact stars even if it is not absolutely stable. 
The existence of absolutely stable strange quark matter is still 
an open question and it may be realized in the form of 
strangelets (small lumps of strange quark matter 
with roughly the same amount of up, down and strange quarks). 
Theoretically, there has been various investigations 
of the stability of (non)strange quark matter. 
Within the MIT bag model, Farhi and Jaffe \cite{FJ} found a reasonable window 
of the model parameters (i.e. the bag constant, current quark masses and 
strong coupling constant) for which 
strange quark matter is stable, while nonstrange quark matter 
is unstable as compared to a gas of $^{56}\mbox{Fe}$. 
By including finite size effects (surface tension and/or curvature energy), 
similar analyses have been done 
for finite lumps of quark matter.\cite{SQM,FJ,BJ,GJ,JM} 
It is now well established that finite size effects 
increase the energy of finite quark lumps.

During the last decade, significant advances have been made 
in our understandings of the phase structure of 
hot and/or dense QCD. \cite{CSC} 
At the present time, it is widely accepted that a color-flavor locked 
(CFL) phase \cite{Alford98} is the ground state of cold, dense quark matter.
\footnote{As discussed in Refs. \cite{IB,AR,SRP,NBO}, 
it is of importance to take account of the constraint 
imposed by color and electric neutrality. 
For neutrino free, color and electrically neutral 
quark matter at zero temperature, 
it has been shown that the CFL phase optimizes 
the pairing energy and, then, 
is favored over the 2-flavor superconducting (2SC) phase 
in all (or almost all) the range of densities.\cite{AR,SRP}} 
Madsen \cite{CFLS} has studied the stability of CFL strangelets 
and found that CFL strangelets are significantly more stable than 
normal (unpaired) strangelets (see also Ref. \cite{Lugones1}). 
However, that work ignored issues of density dependence 
of the bag constant and the CFL gap. 
What remains a question is the possible effects of 
the phase structure. 
In the previous paper \cite{KH}, 
we have studied the chiral and 2SC phase of finite quark lumps. 
We found that finite size effects enhance the restoration of 
chiral symmetry. In this case, it is likely that strangelets 
lie in the color superconducting phase.

The purpose of this work is to study 
the behavior of the CFL gap in strangelets and 
to look at its effects on the stability of CFL strangelets. 
In order to describe the density dependent bag constant 
and the density dependent CFL gap we choose to use 
the three-flavor Nambu--Jona-Lasinio (NJL) model.\cite{NJ} 
(The NJL model is a simple tractable model 
to investigate the stability of quark matter/droplets 
[see, for example, Ref. \cite{eos}].) 
We consider up, down and strange quarks to be massless, for simplicity. 
In order to take account of finite size effects, 
we apply what is called multiple reflection expansion (MRE).\cite{MRE} 
The MRE has been used for calculating thermodynamic quantities 
such as the energy per baryon number and the free energy 
of finite quark lumps. As far as the general structure is concerned, 
the results are in good agreement with those given 
in the MIT bag model.\cite{SQM,JM} 
Using the NJL model with the MRE, we formulate 
the thermodynamic potential for spherical CFL strangelets. 
We then investigate the stability of CFL strangelets, 
including the dynamical effects as well as the finite size effects.

The organization of this paper is as follows. 
In the next section, we briefly review 
the thermodynamic potential of the CFL phase in bulk. 
Then, we formulate the thermodynamic potential 
for CFL strangelets. The gap equation and 
related thermodynamic quantities are also derived. 
In Sec. 3, we present numerical results. 
Finally, Sec. 4 is devoted to the conclusions.

\section{Thermodynamic potential} 
First, we briefly survey the thermodynamic potential 
of the CFL phase in bulk. 
To describe the CFL phase in the NJL model 
we follow Buballa and Oertel.\cite{BO2} 
The $\mbox{U}(3)_L \times \mbox{U}(3)_R$ symmetric NJL Lagrangian is
\begin{eqnarray}
{\cal L}=\bar{q}i\gamma^\mu\partial_\mu q
+G_1\sum_{i=0}^{8}\left[(\sigmac)^2+(\pic)^2\right],
\label{eqn:lag}
\end{eqnarray}
where $q$ denotes a quark field with three flavors ($N_f=3$) 
and three colors ($N_c=3$), the coupling constant 
$G_1$ has a dimension $[G_1]=[\mbox{mass}]^{-2}$ and the Gell-Mann matrices 
$\tau^{i}$ $(i=1,\cdots,8)$ with $\tau^0=\sqrt{2/3}\bvec{I}_f$ act 
in the flavor space. 

In the case of the CFL, the quarks form 
the following $(\bar{3}_c,\bar{3}_f)$ condensate
\begin{eqnarray}
\Delta_{ij}^{\alpha\beta} 
=\langle q_{i}^{\alpha}q_{j}^{\beta}\rangle
\propto C\gamma_5\epsilon^{\alpha\beta X}\epsilon_{ijX},
\end{eqnarray}
where $i,j$ denote flavor indices, $\alpha, \beta$ denote 
color indices and repeated indices are summed. 
In principle, there must be the additional $(6_c,6_f)$ condensate. 
However, we neglect such condensate because it is much smaller 
than $(\bar{3}_c,\bar{3}_f)$ condensate \cite{Alford98,Shovkovy99} 
and its contribution to the thermodynamic potential is negligible. 
The interaction corresponding to 
the $(\bar{3}_c,\bar{3}_f)$ CFL paring can be written as
\begin{eqnarray}
{\cal L}_{\mb{CFL}}
=G_2\sum_{i=\alpha=2,5,7}
\left(\bar{q}i\gamma_5\tau^{i}\lambda^{\alpha}C\bar{q}^T\right)
\left(q Ci\gamma_5\tau^{i}\lambda^{\alpha}q\right),
\end{eqnarray}
where $C$ is a charge conjugation matrix, 
defined by $C^{-1}\gamma_{\mu}C=-\gamma_{\mu}^T$ and $C^T=-C$, and 
$\tau^i$ $(i=2,5,7)$ [$\lambda^{\alpha}$ $(\alpha=2,5,7)$] denote 
the antisymmetric 
generators of $\mbox{SU}(3)_f$ [$\mbox{SU}(3)_c$]. 
The coupling constant $G_2$ can be obtained from Eq. (\ref{eqn:lag}) 
by making use of the Fierz transformation. 
However, we leave $G_2$ a free parameter. 
In this approximation, all nine (three colors times three flavors) 
quarks participate in the pairing 
and yield eight quasiparticles with gap $\Delta$ 
and one with gap $-2\Delta$. 
(These quasiparticles correspond to an octet and a singlet 
of the unbroken $\mbox{SU}(3)$, respectively.) 
Then, in the mean-field approximation, the thermodynamic potential 
$\Omega=\Omega(\Delta;\mu,T)$ at finite quark chemical potential $\mu$ 
and temperature $T$ is given by
\begin{eqnarray}
\Omega&=&\frac{3\Delta^2}{4G_2}
-8T\sum_{\pm}\sum_{n=-\infty}^{\infty}\int\frac{d^3k}{(2\pi)^3}
\ln(\omega_n^2+\epsilon_{\pm}^2)\nonumber\\
&&-T\sum_{\pm}\sum_{n=-\infty}^{\infty}\int\frac{d^3k}{(2\pi)^3}
\ln(\omega_n^2+\xi_{\pm}^2),
\end{eqnarray}
where $\epsilon_{\pm}=\sqrt{(k\pm\mu)^2+\Delta^2}$, 
$\xi_{\pm}=\sqrt{(k\pm\mu)^2+4\Delta^2}$ 
and $\omega_n=(2n+1)\pi T~(n=0,\pm 1,\pm 2,\cdots)$ denote 
the fermionic Matsubara frequencies. 
The summation over $\omega_n$ is straightforward. 
Henceforth, we restrict ourselves to $T=0$. 
The $T \to 0$ limit of $\Omega$ is
\begin{eqnarray}
\Omega=\frac{3\Delta^2}{4G_2}
-\sum_{\pm}\int_0^{\Lambda_{\mb{UV}}}\frac{k^2dk}{2\pi^2}
\left(8\epsilon_{\pm}+\xi_{\pm}\right).
\end{eqnarray}
Here, we have introduced the ultraviolet cutoff $\Lambda_{\mb{UV}}$ 
in the three-dimensional momentum space. 

Now we incorporate finite size effects into the thermodynamic potential. 
To this end, we use the density of states 
derived from the MRE.\cite{MRE,JM,BJ} 
In the MRE framework, the density of states 
for a spherical system is given by $k^2\rhom/(2\pi^2)$, 
where
\begin{eqnarray}
\rhom&=&\rhom(k,m,R)\nonumber\\
&=&1+\frac{6\pi^2}{kR}f_S\left(\frac{k}{m}\right)
+\frac{12\pi^2}{(kR)^2}f_C\left(\frac{k}{m}\right)+\cdots,\label{eqn:dos}
\end{eqnarray}
with $m$ being the Dirac mass of quark and $R$ being the radius of the sphere. 
The second (third) term on the right-hand side represent 
the surface (curvature) contribution to the fermionic density of states. 
The ellipsis implies higher order terms in $1/R$, 
which are neglected throughout. 
In the case of massless fermions, the functions $f_S$ and $f_C$ have 
the following limits,
\begin{eqnarray}
\lim_{m \to 0}f_S(k/m)=0~,~~\lim_{m \to 0}f_C(k/m)=-1/(24\pi^2).
\end{eqnarray}
Then, we use the following MRE density of states,
\begin{eqnarray}
\rhom=1-\frac{1}{2(kR)^2}.\label{eqn:dos2}
\end{eqnarray}
One can see that the finite size effects reduce the density of states 
(the decreasing tendency is more pronounced at low momenta) 
and $\rhom$ becomes negative at small $kR$. 
To avoid the unphysical negative density of states 
we shall introduce an infrared cutoff 
$\Lambda_{\mb{IR}}=\sqrt{2}/(2R)$ in the momentum space.

Using the density of states (\ref{eqn:dos2}), 
we express the effective potential of the spherical CFL strangelets 
as follows,
\begin{eqnarray}
\Omega_{\mb{MRE}}=\frac{3\Delta^2}{4G_2}
-\sum_{\pm}\mre\left(8\epsilon_{\pm}+\xi_{\pm}\right),\label{eqn:ep2}
\end{eqnarray}
where we have introduced the following notation,
\begin{eqnarray}
\mre=\int_{\Lambda_{\mb{IR}}}^{\Lambda_{\mb{UV}}}\frac{k^2dk}{2\pi^2}\rhom.
\end{eqnarray}
Here, we emphasize that the Fermi momentum $k_F(=\mu)$ 
in Eq. (\ref{eqn:ep2}) is common to all nine quarks. 
In the case of the three flavors of massless quarks, 
a strangelet with the common Fermi momentum is automatically 
color and electrically neutral.

A strangelet must be in a color singlet state. 
As noted in Refs. \cite{AR,ABMW}, 
color neutrality is nothing but a prerequisite condition 
for color singletness. 
However, it has been shown that, 
as long as a quark lump of color superconductor is color neutral, 
the projection over a color singlet state makes a negligible 
contribution to the thermodynamic potential.\cite{ABMW} 
Thus, color neutrality is a good approximation 
to color singletness. A small strangelet $(A \ll 10^7)$ 
does not need to satisfy electric neutrality. 
This is because the electron Compton wavelength is larger than  
such a strangelet and, then, electrons mainly stay 
outside of the quark phase.\cite{SQM} 
Therefore the phases of small strangelets could be different from 
that of quark matter in bulk. 

We focus on CFL strangelets embedded in a vacuum and 
derive a set of coupled equations with the MRE. 
For computation of a finite system, we choose a fixed radius $R$. 
First, we derive the gap equation, 
which is the extremum condition of $\Omega_{\mb{MRE}}$ 
with respect to $\Delta$:
\begin{eqnarray}
\frac{\partial\Omega_{\mb{MRE}}}{\partial\Delta}=
\frac{3\Delta}{2G_2}-4\Delta\sum_{\pm}\mre
\left(\frac{2}{\epsilon_{\pm}}+\frac{1}{\xi_{\pm}}\right)=0.\label{eqn:sde}
\end{eqnarray}

In addtion, we need to take account of the pressure balence relation. 
The relative pressure inside strangelet $P_{\mb{MRE}}$ is given by
\begin{eqnarray}
P_{\mb{MRE}}=p_{\mb{MRE}}-p_{\mb{vac}},\label{eqn:pressure}
\end{eqnarray}
where $p_{\mb{MRE}}=-\Omega_{\mb{MRE}}$ and $p_{\mb{vac}}$ has been introduced 
to measure the pressure relative to outside of strangelet. 
Note that $p_{\mb{vac}}$ is nothing but the pressure 
of chirally broken vacuum. 
In the mean-field approximation, its value can be determined 
by using the Lagrangian (\ref{eqn:lag}) as follows:
\begin{eqnarray}
p_{\mb{vac}}=2N_fN_c\int_0^{\Lambda_{\mb{UV}}}
\frac{d^3k}{(2\pi)^3}\sqrt{k^2+M^2}-\frac{3M^2}{8G_1},
\end{eqnarray}
Here, $M$ is the dynamically generated quark mass, 
which is a nontrivial solution to the following equation,
\begin{eqnarray}
\frac{3M}{4G_1}-2N_fN_c\int_0^{\Lambda_{\mb{UV}}}
\frac{d^3k}{(2\pi)^3}\frac{M}{\sqrt{k^2+M^2}}=0.
\end{eqnarray}

We solve Eqs. (\ref{eqn:sde}) and (\ref{eqn:pressure}) self-consistently 
and, then, compute the baryon number of the strangelet $A$ 
by making use of the following relation,
\begin{eqnarray}
A=Vn_B,\label{eqn:bnd}
\end{eqnarray}
where $V=4\pi R^3/3$ is the volume of the spherical strangelet and
\begin{eqnarray}
n_B=-\frac{1}{3}
\frac{\partial\Omega_{\mb{MRE}}}{\partial\mu}
=\frac{1}{3}\sum_{\pm}\mre
\left[\mp\frac{8(k\pm\mu)}{\epsilon_{\pm}}\mp\frac{k\pm\mu}{\xi_{\pm}}\right],
\end{eqnarray}
is the baryon number density of the strangelet.

\section{Numerical results}
%
%
\begin{figure}
\includegraphics[scale=0.7]{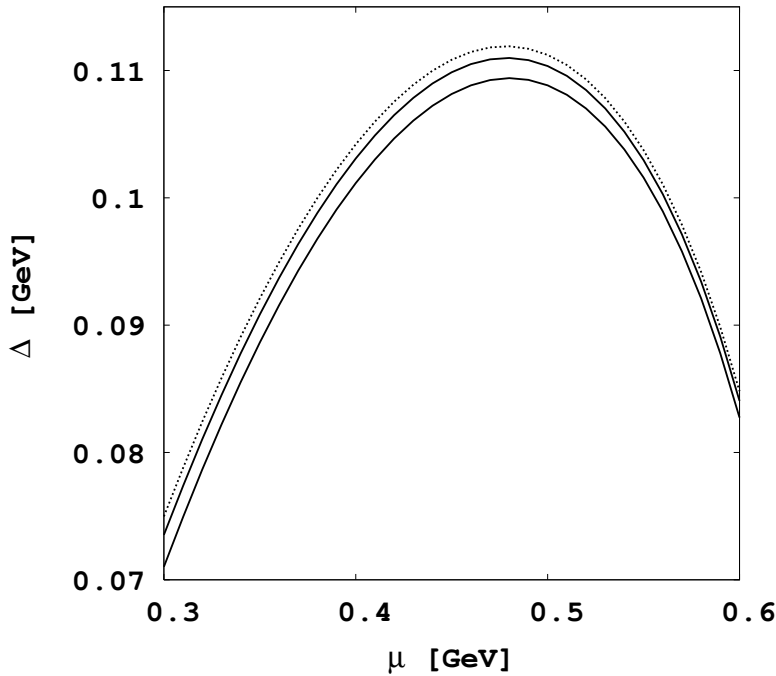}
\caption{The CFL gap $\Delta$ for finite volumes. 
The solid curves are obtained for three values of $R$: 
$R=3,5$ fm (from bottom to top). 
The dotted line refers to the case of $R\to\infty$. 
We have taken $G_2/G_1=3/4$.}
\end{figure}
%
%
\begin{figure}
\includegraphics[scale=0.7]{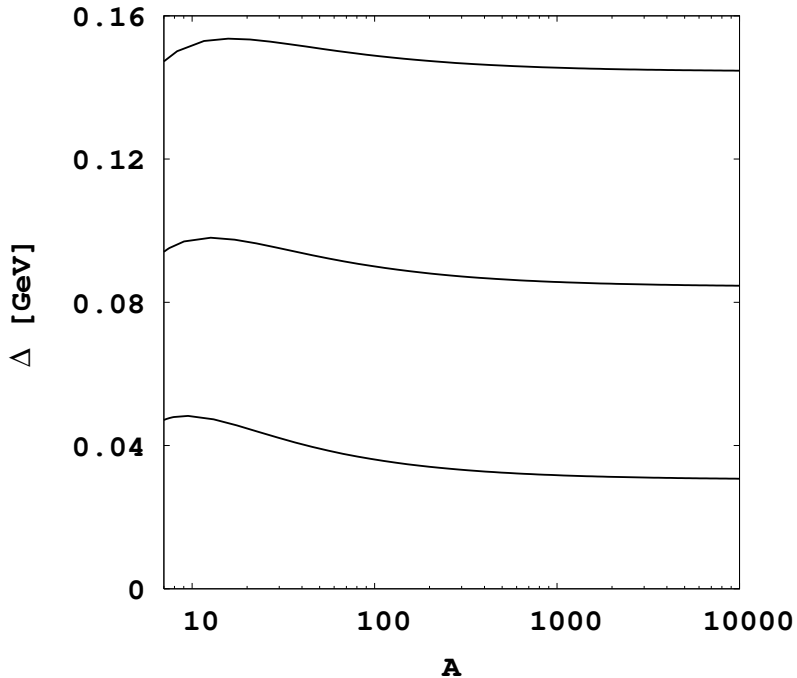}
\caption{The CFL gap $\Delta$ as functions of the baryon number $A$. 
The curves correspond to the cases of $G_2/G_1=1/2,3/4,1$ 
(from bottom to top).}
\end{figure}

In numerical calculation, we use the following set of parameters: 
$\Lambda_{\mb{UV}}=0.6$ GeV, $G_1\Lambda^2=2.311$.\cite{BO2} 
At $T=\mu=0$, these parameters yield 
the constituent mass $M=0.35$ GeV 
and the bag constant $B^{1/4}=0.182$ GeV. 
The quark-quark coupling constant $G_2$ 
has been set to be $G_2/G_1=3/4$ in the literature.\cite{BO2,SRP,GNA} 
This relation between $G_1$ and $G_2$ is obtained from a four-fermion 
interaction with a quantum number of a one-gluon exchange, 
\begin{eqnarray}
{\cal L}_{\mb{OGE}}=-g\sum_{\alpha=1}^8
(\bar{q}\gamma^{\mu}\lambda^{\alpha}q)^2,
\end{eqnarray}
by making use of the Firez transformation. 
Moreover, this is consistent with that determined 
by fitting the nucleon mass within a Fadeev approach.\cite{BT} 
However, we vary $G_2$ in the range of 
$G_2/G_1=(1/2-1)$ to see its effect 
on the stability of strangelets.

We begin with the discussion of the solution to 
the gap equation (\ref{eqn:sde}). 
Figure 1 shows the CFL gap for finite volumes 
as a function of the quark chemical potential. 
We find that the finite size effects reduce the gap. 
However, the decrease in the gap is not pronounced even at small radii; 
therefore, the $R\to\infty$ limit is a good approximation 
as to the size of the gap.
\footnote{In this respect the result is similar to 
that in Ref. \cite{ABMW} where finite size effects on the 2SC gap 
have been studied by confining the quarks to a cubic box.} 
This is easily understood as follows. 
Color superconductivity is brought about by the quarks 
near the Fermi surface. 
Recall that at a fixed radius $R$ 
the decrease in the density of states 
is not pronounced at large momenta. 
Then, if the Fermi momentum is sufficiently large, 
the finite size effects do not greatly affect the gap.

In order to look at CFL strangelets, we solve 
the gap equation (\ref{eqn:sde}), 
the pressure balance relation (\ref{eqn:pressure}) 
and the baryon number relation (\ref{eqn:bnd}), 
self-consistently. 
By numerically solving these equations, 
we obtain the baryon number dependence of the gap, 
the quark chemical potential and the radius. 
After that the energy per baryon number 
of the pressure-balanced strangelet,
\begin{eqnarray}
\frac{E}{A}\bigg{|}_{p=0}=\frac{{\cal E}}{n_B}\bigg{|}_{p=0}=3\mu,
\label{eqn:epb}
\end{eqnarray}
can be evaluated. 
In Eq. (\ref{eqn:epb}), 
${\cal E}=\Omega_{\mb{MRE}}+\mu\sum_{a}n_a-\epsilon_{\mb{vac}}$ denotes 
the energy density of the strangelet, where $a\in(u,d,s)$, 
$n_u=n_d=n_s(=n_B)$ are the quark number densities, 
and $\epsilon_{\mb{vac}}(=-p_{\mb{vac}})$ is the energy density 
of the chirally broken vacuum.

Figure 2 shows the CFL gap as a function of 
the baryon number for the cases of $G_2/G_1=1/2,3/4,1$. 
Each curve remains approximately constant 
as long as $A$ is not too small. 
The change of the gap at small baryon numbers $(A \lessapprox 100)$ 
is a consequence of the fact that the actual quark chemical potential 
increases when $A$ decreases. 
We note that the results for $A \lessapprox 10$ should not be considered 
robust ones, because in this regime 
the quark chemical potentials are close 
to the ultraviolet cutoff $\Lambda_{\mb{UV}}$.

%
%
\begin{figure}
\includegraphics[scale=0.7]{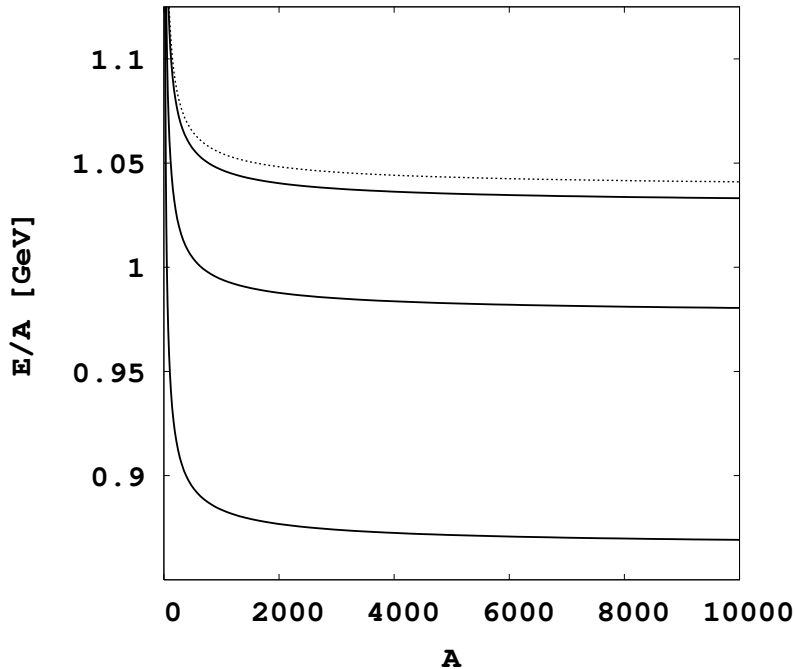}
\caption{The energy per baryon $E/A$ for CFL strangelets. 
The solid curves correspond to the cases of 
$G_2/G_1=1/2,3/4,1$ (from top to bottom). 
The dotted line refers to $E/A$ for normal strangelets.}
\end{figure}

In Fig. 3, the energy per baryon number $E/A$ for CFL strangelets 
is plotted as a function of the baryon number. 
For comparison, we also present the result for normal strangelets. 
As is evident from the figure, 
CFL strangelets are considerably more stable than normal strangelets. 
Needless to say, this decrease in $E/A$ is due to 
the pairing energy contribution 
to the thermodynamic potential $\Omega_{\mb{MRE}}$. 
Note that $E/A$ depends on 
the size of the gap; hence, the quark-quark coupling constant $G_2$. 
As $G_2$ is increased, the gap increases and then 
the contribution to $\Omega_{\mb{MRE}}$ 
from the pairing energy also increases. 
Thus, it is reasonable that the growth of $G_2$ 
tends to lower $E/A$ of CFL strangelets. 
The gap can be large enough to have a great effect on $E/A$. 
For instance, we observe that at $G_2=G_1$ 
CFL strangelets of $A \ge 160$ are absolutely stable. 
Within our model parameters ($\Lambda_{\mb{UV}}$ and $G_1$), 
$G_2/G_1$ needs to exceed $0.88$ for the existence 
of the absolutely stable CFL strangelets. 
Note also that each curve shows the typical behavior 
of $E/A$ of finite quark lumps. 
$E/A(=3\mu)$ increases with decreasing $A$. 
(As mentioned earlier, this is a consequence of the increase of 
the quark chemical potential by the finite size effects.) 
On the other hand, as $A$ grows, $E/A$ approaches the value 
without the finite size effects. 

\section{Conclusions} 
In this work, we have explored the color-flavor locked phase in strangelets. 
We used the three-flavor Nambu--Jona-Lasinio model. 
To take account of finite size effects we applied 
the multiple reflection expansion. 
Using the NJL model with the MRE, we formulated 
the thermodynamic potential for CFL strangelets 
and derived related thermodynamic quantities. 
We clarified the behavior of the CFL gap 
and its effects on the stability of strangelets, 
although our model is nothing but a toy model, 
neglecting the finite strange quark mass. 
The CFL gap is almost independent of $A$ as long as $A$ is not too small. 
Due to the contribution from the formation of the CFL condensate, 
CFL strangelets are more stable than strangelets without the CFL. 
Further, if $G_2$ is larger than the critical value, 
we have a chance to find absolutely stable CFL strangerets. 
These results complement some of the conclusions made by Madsen.\cite{CFLS} 
Of course we cannot completely accept the existence of 
absolutely stable CFL strangelets, 
because we do not have a good knowledge of $G_2$. 
However, a Fadeev approach to baryons \cite{BT,BHIT} would be a guide 
to the determination of $G_2$. 
Moreover, we had neglected many aspects of dense QCD. 
Most important is probably the finite strange quark mass $m_s$. 
When $m_s$ is nonzero, we have to take account of 
unlocking phase transition.\cite{Alford99} 
As pointed out in Ref. \cite{NBO}, 
the main reason for the difficulty to find the stable 2SC matter 
arises from the constraint of electric neutrality. 
In contrast, small strangelets, which can be regarded 
as free from this constraint, could be in the 2SC phase.\cite{Kiriyama} 
It should be mentioned, however, that 
the increase of the chemical potential 
caused by the finite size effects might prefer the CFL phase 
to the 2SC phase. 
It would be of great importance to study competition 
with other phases (hadronic phase, 2SC phase, and so on), 
taking account of nonzero $m_s$ 
as well as color neutrality.\cite{IB,AR,SRP,NBO} 
A more careful analysis is left to future studies.

It should be also noted that the MRE contains several problems 
concerning its reliability. 
First, the density of state $\rhom$ should have the terms 
proportional to $1/R^3$, $1/R^4$, and so on. 
These terms would be dominant at small radii, i.e., 
at small baryon numbers. 
Furthermore, $\rhom$ in Eq. (\ref{eqn:dos}) causes 
the unphysical negative density of states at small $kR$. 
Although our qualitative results for relatively 
large baryon numbers $(A \gtrapprox 100)$ would not change, 
it is desirable to take more rigorous way 
of including finite size effects. 
In particular, as noted in Ref. \cite{JM}, a detailed study of 
strangelet decay modes will have to rely on shell model calculations. 

In this work, we restricted ourselves to zero temperature. 
However, for applications such as 
heavy ion collision experiments and the cosmological 
quark-hadron phase transition, 
it is interesting to investigate strangelets at finite temperature. 
The present approach is straightforwardly generalized to 
nonzero temperature. Studies along this line are now in progress.

\end{document}